\documentclass[journal=jctcce,keywords=true,manuscript=article]{achemso}

\usepackage[english]{babel}
\usepackage[utf8x]{inputenc}
\usepackage[T1]{fontenc}
\usepackage{comment}

\usepackage{amsmath}
\usepackage{amsthm}
\usepackage{color}
\usepackage{caption}
\usepackage{subcaption}
\usepackage{xcolor}
\usepackage{amsfonts}
\usepackage{graphicx}
\usepackage{multirow}
\usepackage{mathtools}
\usepackage[colorinlistoftodos]{todonotes}
\usepackage[colorlinks=true,allcolors=blue]{hyperref}

\usepackage[normalem]{ulem}

\DeclarePairedDelimiter\ceil{\lceil}{\rceil}

\author{Pengfei Li}
\affiliation{Silicon Therapeutics, Suzhou, Jiangsu 215000, China}
\author{Zhijie Li}
\affiliation{Silicon Therapeutics, Suzhou, Jiangsu 215000, China}
\author{Yu Wang}
\affiliation{Silicon Therapeutics, Suzhou, Jiangsu 215000, China}
\author{Huaixia Dou}
\affiliation{Silicon Therapeutics, Suzhou, Jiangsu 215000, China}
\author{Brian K. Radak}
\affiliation{Roivant Sciences, Boston, Massachusetts 02210, United States}
\author{Woody Sherman}
\affiliation{Roivant Sciences, Boston, Massachusetts 02210, United States}
\author{Huafeng Xu}
\email{huafeng@gmail.com}
\affiliation{Roivant Sciences, New York, New York 10036, United States}

\title{Precise binding free energy calculations for multiple molecules using an optimal measurement network of pairwise differences}

\begin{document}
\maketitle

\begin{abstract}
Alchemical binding free energy (BFE) calculations offer an efficient and thermodynamically rigorous approach to {\it in silico} binding affinity predictions.  As a result of decades of methodological improvements and recent advances in computer technology, alchemical BFE calculations are now widely used in drug discovery research.  They help guide the prioritization of candidate drug molecules by predicting their binding affinities for a biomolecular target of interest (and potentially selectivity against undesirable anti-targets).  Statistical variance associated with such calculations, however, may undermine the reliability of their predictions, introducing uncertainty both in ranking candidate molecules and in benchmarking their predictive accuracy.  Here, we present a computational method that substantially improves the statistical precision in BFE calculations for a set of ligands binding to a common receptor by dynamically allocating computational resources to different BFE calculations according to an optimality objective established in a previous work from our group and extended in this work. Our method, termed Network Binding Free Energy (NetBFE), performs adaptive binding free energy calculations in iterations, re-optimizing the allocations in each iteration based on the statistical variances estimated from previous iterations.  Using examples of NetBFE calculations for protein-binding of congeneric ligand series, we demonstrate that NetBFE approaches the optimal allocation in a small number ($\leq 5$) of iterations and that NetBFE reduces the statistical variance in the binding free energy estimates by approximately a factor of two when compared to a previously published and widely used allocation method at the same total computational cost.
\end{abstract}
\clearpage

\section{Introduction}
In drug discovery research, candidate drug molecules are increasingly ranked and selected by their computationally predicted affinities for the biomolecular target of interest before experimental synthesis and testing~\cite{Chodera2011COSB,Mobley2012JCP,Abel2017ACR,Schindler2020JCIM,Song2020JCIM,Cournia2020JCIM}.  This allows for the exploration of many more molecular design ideas than is practical with experimental approaches alone.  Alchemical binding free energy (BFE) calculations have become a valuable tool in predicting the protein-ligand binding affinities and their utility continues to increase thanks to advances in the methodologies~\cite{Mey2020LJCMS,Tsai2020JCIM,Lee2020JCIM,He2020ACSOmega,Wang2015JACS, Shih2020JCIM,Schindler2020JCIM,Chatterjee2017JACS} force field~\cite{Roos2019JCTC,Lu2021JCTC,Vanommeslaeghe2010JCC, Wang2004JCC,Halgren1996JCC,Horta2016JCTC}, computational power afforded by the graphical prcessing units (GPUs)~\cite{Wang2015JACS,Lee2020JCIM,Tsai2020JCIM}, and the availability of simulation software and simplified workflows~\cite{Wang2015JACS,Lee2020JCIM,Schindler2020JCIM,Christ2014JCIM, Loeffler2015JCIM, Fu2018JCIM, Lundborg2015JPCB, Klimovich2015JCAMD, Gapsys2015JCC,Senapathi2020JCIM}.

BFE calculations can either compute the binding free energy of an individual molecule, by a technique commonly referred to as "absolute" binding free energy (ABFE) calculations~\cite{Boresch2003JPCB,Mobley2007JMB,Woo2005PNAS,Aldeghi2016CS}, or compute the difference between the binding free energies of two molecules, by "relative" binding free energy (RBFE) calculations~\cite{Reddy1994JMC,Reddy2001JACS,Wang2015JACS,Cournia2017JCIM}.  The binding free energies of a set of molecules can be determined by a combination of ABFE calculations of select molecules and RBFE calculations of select pairs of molecules~\cite{Xu2019JCIM}.

Statistical errors in BFE calculations confound  decision-making in candidate molecule selection and the assessment of their predictive accuracy.  There have been significant methodological progress in minimizing statistical errors in single BFE calculations~\cite{Shirts2008JCP,Shenfeld2009PRE}, but works have only recently been published addressing the question of how to reduce the collective statistical errors for BFE calculations of a set of molecules~\cite{Xu2019JCIM,Yang2020JCC}, which represents the majority of applications of BFE calculations in drug discovery.

As suggested previously, at a fixed total computational cost, the overall statistical errors in BFE calculations for multiple molecules can be minimized by optimally allocating the computational resources to different RBFE and ABFE calculations.  In particular, minimizing the total variance in the estimated binding free energies with respect to the allocation corresponds to the A-optimal in experimental design~\cite{Xu2019JCIM}.  Such optimal allocations, however, require knowledge of the statistical fluctuations in individual BFE calculations, which are unknown {\it a priori}.  Because of this reason, A-optimal BFE calculations have not yet been reported for realistic examples. 

In this work, we develop a method based on the A-optimal design to improve the statistical precision of BFE calculations.  Our method, Network Binding Free Energy (NetBFE), performs adaptive BFE calculations in multiple iterations.  In each iteration, NetBFE estimates the statistical fluctuations in different BFE calculations based on the results from previous iterations, and it reallocates the computational resources accordingly.  This allows the A-optimal allocation to be used in realistic BFE calculations.  We also extend the previous theoretical work by deriving new equations for the A-optimal design and the covariance matrix when only RBFE calculations are included (the previous theory required the inclusion of at least one ABFE calculation), a common scenario in the lead optimization stage of drug discovery.  

Here, we apply NetBFE to calculating the binding free energies of 16 ligands against the TYK2 protein~\cite{Wang2015JACS,Liang2013JMC} and 8 ligands against the HIF-2$\alpha$ protein~\cite{Scheuermann2015JMC}.  We compare the statistical errors in the estimated binding free energies from NetBFE calculations and those in the BFE calculations using RBFE pairs selected by the LOMAP algorithm developed by the Mobley lab~\cite{Liu2013JCAMD}, which is widely used in such calculations.  At the same total computational cost, NetBFE reduces the statistical variance by about half compared to LOMAP.  Our results suggest that NetBFE can substantially improve the statistical precision of BFE calculations in drug discovery, which should accelerate the computational selection of candidate drug molecules and thereby increase the efficiency of the design-make-test cycle.

\section{Methods}
\subsection{A-optimal design of network binding free energy calculations}
The A-optimal design of measurement network of pairwise differences was described previously in detail~\cite{Xu2019JCIM}, and is briefly explained here.  The binding free energy values for multiple molecules in the same binding site of a receptor are the quantities of interest in BFE calculations. Let $x_{i=1,2, \ldots, m}$ be the binding free energy of molecule $i$; here we use $x_i$ instead of $\Delta G_i$ to make the notations in the mathematical presentation more compact, and to follow the same notation in Reference \citenum{Xu2019JCIM}.  ABFE computes the individual $x_i$ ({\it i.e.} $\Delta G_i$) and RBFE computes the difference between any pair of $x_{i}$ and $x_{j}$: $x_{ij} = x_i - x_j$ ({\it i.e.} $\Delta\Delta G_{ij} = \Delta G_i - \Delta G_j$).  These ABFE and RBFE calculations form our difference network, which can be represented by a graph $\mathcal{G}$ of $m+1$ vertices and $m(m+1) / 2$ edges, where the vertices $i=1,2, \ldots, m$ stand for the $m$ quantities $\left\{x_{i}\right\}$, the edge between vertices $i \neq j>0$ stands for the difference measurement $\hat{x}_{i j}$ ({\it i.e.} the RBFE calculations), and the edge between vertices 0 and $i>0$ stands for the individual measurement $\hat{x}_{i}$ ({\it i.e.} the ABFE calculations). For each BFE calculation, its asymptotic statistical variance $\sigma_{e}^{2}$, for $e \in\{i|i=1,2, \ldots, m\} \cup \{(i, j) \mid i, j=1,2, . ., m, i \neq j\}$, decreases with $n_{e}$---the resource allocated to this calculation---as
\begin{equation}
\sigma_{e}^{2}=s_{e}^{2} / n_{e}
\label{eq:variance-s-n}
\end{equation}
where $s_{e}$ is the statistical fluctuation in the corresponding BFE calculation.

Some of the binding free energy values, say $\left\{x_{a} \mid a \in Q\right\}$, may have been measured experimentally.  These experimental values and their concomitant statistical errors in the experiment, $\left\{\tilde{x}_{a} \pm \delta_{a} \mid a \in Q\right\}$, can be included in the estimate of the binding free energies from BFE calculations. If the true binding free energies are $\{ x_i \}$, the likelihood that our BFE calculations produce results of $\{\hat{x}_e \pm \sigma_e \}$ and the experimental measurements produce results of $\{\tilde{x}_a \pm \delta_a \mid a \in Q \}$ is 
\begin{eqnarray}
L &=& \prod_i (\sqrt{2\pi \sigma_i})^{-1} 
  \exp\left( -\frac{(x_i - \hat{x}_i)^2}{2\sigma_i^2} \right)
\nonumber \\
  &\cdot& \prod_{a\in Q} (\sqrt{2\pi \delta_a})^{-1}
  \exp\left( -\frac{(x_a - \tilde{x}_a)^2}{2\delta_a^2} \right)
\nonumber \\
  &\cdot& \prod_{i<j} (\sqrt{2\pi \sigma_{ij}})^{-1}
  \exp\left( -\frac{(x_i - x_j - \hat{x}_{ij})^2}{2\sigma_{ij}^2}\right)
\label{eq:likelihood}
\end{eqnarray}

Maximizing the log-likelihood $\ln L$ with respect to $\{ x_i \}$ yields the maximum likelihood estimator for $\left\{x_{i}\right\}$:
\begin{equation}
\mathbf{F} \cdot \vec{x}=\vec{z}
\label{eq:MLE}
\end{equation}
where $\mathbf{F}$ is the Fisher information matrix with the elements
\begin{equation}
F_{i j}=\left\{\begin{array}{cl}
\delta_{i}^{-2}+\sigma_{i}^{-2}+\sum_{k \neq i} \sigma_{i k}^{-2} & \text {if } i=j \text { and } i \in Q \\
\sigma_{i}^{-2}+\sum_{k \neq i} \sigma_{i k}^{-2} & \text {if } i=j \text { and } i \notin Q \\
-\sigma_{i j}^{-2} & \text {if } i \neq j
\end{array}\right.
\label{eq:Fisher}
\end{equation}
and
\begin{equation}
z_{i}=\left\{\begin{array}{cl}
\delta_{i}^{-2} \tilde{x}_{i}+\sigma_{i}^{-2} \hat{x}_{i}+\sum_{j \neq i} \sigma_{i j}^{-2} \hat{x}_{i j} & \text { if } i \in Q \\
\sigma_{i}^{-2} \hat{x}_{i}+\sum_{j \neq i} \sigma_{i j}^{-2} \hat{x}_{i j} & \text {if } i \notin Q.
\end{array}\right.
\label{eq:z}
\end{equation}
The covariance in the estimates of $\left\{x_{i}\right\}$ is given by the inverse of the Fisher information matrix:
\begin{equation}
\mathbf{C}=\mathbf{F}^{-1}.
\end{equation}
The A-optimal minimizes the total variance, tr($\mathbf{C}$), subject to the constraints of non-negativity
\begin{equation}
n_{e} \geq 0,
\end{equation}
and of the total fixed computational cost
\begin{equation}
\sum_{e} n_{e}=\sum_{i} n_{i}+\sum_{i<j} n_{i j} = N.
\end{equation}
This minimization can be solved by cone programming\cite{Stephen2004CUP}, and it is implemented in the publicly available software DiffNet.~\cite{XuDiffNet}%\href{https://github.com/forcefield/DiffNet}{https://github.com/forcefield/DiffNet}.

\subsection{A-optimal for a network with only relative binding free energy calculations}
Sometimes we may choose to perform only relative binding free energy calculations and compute only $\hat{x}_{ij}$.  In this case the free energies can only be determined up to an arbitrary constant $\bar{x}$, which can be chosen to minimize the root mean square deviation from the known experimental values. 

The log-likelihood in this case is given by
\begin{equation}
    \ln L = -\sum_{i<j} \frac{(x_i - x_j - \hat{x}_{ij})^2}{2\sigma_{ij}^2}
\label{eq:lnP-RBFE-only}
\end{equation}

The corresponding Fisher information matrix
\begin{equation}
    F_{ij} = \left\{
    \begin{array}{cl}
    \sum_{j\neq i} \sigma_{ij}^{-2} & \text{ if } i=j \\
    -\sigma_{ij}^{-2} & \text{ if } i\neq j
    \end{array}
    \right.
\end{equation}
is singular. Consequently, the covariance matrix $\mathbf{C}=\mathbf{F}^{-1}$ cannot be defined.

This problem can be solved by introducing an additional term to the log-likelihood to restrain the mean $m^{-1} \sum_i x_i$ to a constant $\bar{x}$:
\begin{equation}
    \ln L^\ast(w) = \ln L - 2^{-1}\omega \left( m^{-1}\sum_i x_i - \bar{x} \right)^2
\end{equation}
which is equivalent to specifying that the mean of $\{ x_i \}$ is normally distributed around $\bar{x}$ with the standard deviation of $1/\sqrt{\omega}$.  The corresponding Fisher information matrix is
\begin{equation}
\mathbf{F}^\ast(w) = \mathbf{F} + \omega m^{-2} \mathbf{1}
\end{equation}
where $\mathbf{1}$ is a $m\times m$ matrix of elements 1.  We substitute this augmented Fisher matrix $\mathbf{F}^\ast(\omega)$ for $\mathbf{F}$ in the semi-definite programming for A-optimal when there are no absolute binding free energy calculations.  It can be demonstrated that the optimal allocation does not depend on the value of $\omega > 0$.

The corresponding covariance matrix is
\begin{equation}
\mathbf{C}^\ast(\omega) = (\mathbf{F}^\ast(\omega))^{-1}
\end{equation}

We are interested in $\mathbf{C}^\ast = \lim_{\omega\rightarrow \infty}\mathbf{C}^\ast(\omega)$, such that the mean of $\{ x_i \}$ is constrained to the constant $\bar{x}$.  To derive $\mathbf{C}^\ast$, consider the expansion of $\mathbf{C}^\ast(\omega)$ in $\omega^{-1}$:
\begin{equation}
    \mathbf{C}^\ast(\omega) = \mathbf{C}^\ast + \omega^{-1} \mathbf{C}_1 + o(\omega^{-2}).
\end{equation}

We have
\begin{eqnarray}
    \mathbf{I} &=& (\mathbf{F} + \omega m^{-2} \mathbf{1})(\mathbf{C}^\ast + \omega^{-1}\mathbf{C}_1 + o(\omega^{-2}))
    \nonumber \\
    &=& \mathbf{F}\mathbf{C}^\ast + m^{-2}\mathbf{1}\mathbf{C}_1
    + \omega m^{-2}\mathbf{1}\mathbf{C} + \omega^{-1}\mathbf{F}\mathbf{C}_1 + o(\omega^{-2}).
\end{eqnarray}

Comparing the terms grouped by different orders of $\omega$, we have
\begin{eqnarray}
\vec{1}^t\mathbf{C}^\ast &=& \vec{0} 
\nonumber \\
\mathbf{F}\mathbf{C}^\ast + \vec{1}m^{-2}\vec{C}_1^t &=& \mathbf{I},
\end{eqnarray}
where $\vec{C}_1^t$ is a vector whose elements are the column sums of $\mathbf{C}_1$.  This can be written in the matrix form
\begin{equation}
    \left( 
    \begin{array}{cc}
    \vec{1}^t & 0 \\
    \mathbf{F} & \vec{1}
    \end{array} 
    \right) \left(
    \begin{array}{c}
         \mathbf{C}^\ast  \\
         m^{-2}\vec{C}_1^t
    \end{array} \right) = \left(
    \begin{array}{c}
         \vec{0}^t  \\
         \mathbf{I}
    \end{array}
    \right),
    \label{eq:C-singular}
\end{equation}
the solution of which gives a well-defined covariance matrix $\mathbf{C}^\ast$.  Incidentally, $\vec{C}_1^t = m \vec{1}^t$.

\subsection{Iterative network binding free energy calculations}
\label{sec:iterative}
The statistical fluctuations $\{ s_e \}$ depend on both the thermodynamic length between the two end states in an alchemical calculation and the ratio of the relaxation time of relevant motions to the length of the simulation.\cite{Shenfeld2009PRE,Mey2020LJCMS} In practice, they have to be estimated (approximately) from the completed BFE calculations.  In order to obtain $\{ s_e \}$ values for allocating the computational resources according to the A-optimal, NetBFE calculations proceed in iterations, with each iteration consuming a predetermined amount, $\Delta N^{(i)}$, of total sampling.  After each iteration $i$, estimates of $\{ s_e^{(i+1)} \} $ are updated as described in Section.~\ref{sec:estimate-se} below, and a new A-optimal allocation $\{ n_e^{(i+1)} \}$ is determined from the updated $\{ s_e^{(i+1)} \}$ and used to extend the BFE calculations for iteration $i+1$. The iterations continue until a specified total simulation samples, $N = \sum_i \Delta N^{(i)}$, or a specified target average statistical variance tr($\mathbf{C}$)/m is reached.  Data accumulated from all the iterations are included to compute the binding free energy value for each BFE calculation.

In each NetBFE calculation reported in this work, a total of $T_0 = m\times 200 \mathrm{ns}$ simulation time is used for a set of $m$ molecules.  This is divided into $I=5$ iterations and in each iteration $T_i = T_0/I$ of simulation time is allocated. 

\subsection{A-optimal on sparse 2-connected sub-networks}
\label{sec:subgraph}
In order to reduce the number of BFE calculations in each iteration, we select a 2-connected subgraph out of the full graph $\mathcal{G}$ in each iteration, and obtain the A-optimal on this sub-network.  It was demonstrated previously that the 2-connected sub-network can generate near-optimal allocations.\cite{Xu2019JCIM}  The allocation in each iteration thus proceeds in two steps in our NetBFE implementation.  First, we determine the A-optimal allocation of the computational resources for the next iteration on the full graph $\mathcal{G}$.  The edges with the largest allocations are selected to form the 2-connected subgraph.  Here, the edge selection can be based on either the total cumulative allocation including all previous iterations or only the allocation for this iteration.  We have implemented both selection methods, and we refer to NetBFE calculations with the former method as NetBFE($N$) and those with the latter method as NetBFE($\Delta N$).  Second, A-optimal allocation is obtained including only the BFE calculations in the selected 2-connected subgraph.  

\subsection{Estimates of the statistical variances in BFE calculations}
The statistical variance $\sigma_{ij}^2$ in the computed $\Delta\Delta G_{ij}$ (by RBFE for $i\neq j$ and by ABFE for $i=j$) is estimated by block bootstrapping, using all the data up to the current iteration. The sampled data points in free energy simulations are divided into $B=10$ non-overlapping blocks, each block containing samples from a simulation time interval. The data are resampled by randomly drawing $B$ blocks with replacement. A $\Delta\Delta G_{ij}$ value is computed from each set of resampled data, and the corresponding variance $\sigma_{ij}^2$ is computed from $K=50$ repeats of the resampling.

\subsection{Estimates of the statistical fluctuations in BFE calculations}
\label{sec:estimate-se}
The statistical fluctuations $s_{ij}$ associated with the calculation of $\Delta\Delta G_{ij}$ can be estimated from its statistical variance $\sigma_{ij}^2$ and its given computational cost $n_{ij}$ by inverting Eq.~\ref{eq:variance-s-n}:
\begin{equation}
    s_{ij}^2 = \sigma_{ij}^2 n_{ij}
\label{eq:s-from-var-and-n}
\end{equation}

We use a simple model to predict $s_{ij}$ of BFE calculations that have not yet been performed (thus $\sigma_{ij}$ is unavailable):

\begin{equation}
    s_{ij} = w_0 + w_1 \max(h_{ij}, h_{ji})^{1/2} + w_2 \max(H_{ij}, H_{ji})^{1/2}
    \label{eq:s-model}
\end{equation}
where for $i\neq j$, $h_{ij}$ is the number of heavy atoms in molecule $i$ that do not map onto atoms in molecule $j$ and $H_{ij}$ is the number of hydrogen atoms in molecule $i$ that do not map onto atoms in molecule j in the RBFE calculation; for $i=j$, $h_{ii}$ and $H_{ii}$ are the respective numbers of heavy and hydrogen atoms in molecule $i$ in the ABFE calculation. 

Before the first iteration, the parameters are initialized to $w_0=1.0$, $w_1=1.0$ and $w_2=0.5$.  In each iteration, the $s_{ij}$ values for the performed RBFE calculations are updated by Eq.~\ref{eq:s-from-var-and-n}, and the parameters $w_0$, $w_1$, and $w_2$ are updated by minimizing the root-mean-square-error (RMSE) between the values predicted from Eq.~\ref{eq:s-model} and those computed from Eq.~\ref{eq:s-from-var-and-n}.

\subsection{Maximum likelihood estimate of binding free energies}
The individual (absolute) binding free energies can be derived from the pairwise (relative) binding free energy differences and their corresponding variances by using the maximum likelihood estimator (Eq.~\ref{eq:MLE}). In practical drug discovery projects, a few molecules usually have experimentally determined binding free energies---reference binding free energies--which can be included in the computational results to derive a maximum likelihood estimate of binding free energies (Eqs.~\ref{eq:Fisher} and~\ref{eq:z}).

\subsection{Molecular dynamic simulations}
In each individual RBFE/ABFE calculation, the protein-ligand complex was first equilibrated by a 5 ns simulation, where the snapshots were saved every 2500 MD steps and the total 2000 snapshots are randomly drawn as the starting structures for replicates of RBFE/ABFE calculations. Hydrogen mass repartition~\cite{HopkinsJCTC2015} in conjunction with SHAKE~\cite{RYCKAERT1977327}/SETTLE~\cite{Miyamoto1992} for constraining the distances between hydrogen atoms and their covalently bonded heavy atoms was employed to enable the use of a large time step of 4 fs in all simulations. The potential energies at all $\lambda$ values were computed and saved every 5 ps. Each alchemical simulation consisted of both the "complex" stage, which simulated the protein-ligand complex, and the "solvated" stage, which simulated the ligand in aqueous solution. %A decoupled protocol (also referred to as "stepwise" transformation)\cite{Tsai2020JCIM, Lee2020JCIM} was used for all BFE calculations in this work. There were 22 $\lambda$ windows in total for RBFE calculation. 
The temperature was kept at 298.15 K using Langevin dynamics~\cite{Niels2013MP} with the collision frequency 2.0 ps$^{-1}$, and the pressure was kept at 1.01325 bar with Monte Carlo barostat~\cite{Aqvist2004CPL} and a pressure relaxation time of 1.0 ps. The proteins were modeled by AMEBR ff14SB force field~\cite{Maier2015JCTC}, and the small molecules were parameterized using our in-house force field generation program; the force field parameters and starting protein-ligand complex structures are provided as AMBER20 input files in the Supplementary Materials. The entire system was solvated in a periodic box of TIP3P water molecules with all the solute atoms no less than 12 {\AA} away from the boundary of the unit cell. In this work, the initial structures of TYK2 protein target with 16 ligands come from reference \citenum{Wang2015JACS} and HIF-2$\alpha$ (PDB ID: 4XT2) protein target with 8 ligands are constructed according to reference\citenum{Scheuermann2015JMC}. All simulations were performed using the AMBER20 package.\cite{AMBER20}

Each individual RBFE calculation consists of $\Lambda = 22$ simulations (decoupled protocol) or $\Lambda = 12$ simulations (concerted protocol) and each individual ABFE calculation consists of $\Lambda = 21$ simulations (decoupled protocol) or $\Lambda = 16$ simulations (concerted protocol) at different $\lambda$ values. Decoupled and concerted protocols are defined in references~\citenum{Song2020JCIM} and~\citenum{Tsai2020JCIM}. If a BFE calculation is allocated a total simulation time of $t_0$, each $\lambda$-valued simulation will run for a total  of $t_\lambda = t_0/\Lambda$ simulation time.  If $t_\lambda < 0.1$ns, this BFE calculation is omitted and its time $t_0$ is allocated proportionally to other BFE calculations, otherwise the simulation time is divided into $\ceil{t_\lambda/(2 \mathrm{ns})}$ replicates initialized with slightly different binding poses and random number seeds (thus no individual simulation exceeds 2 ns simulation time, which helps load balance when the simulations are executed in parallel). In each iteration, new replicates for each individual BFE calculation receiving new allocations are initiated; existing simulations are not extended.  The allocation only includes the "complex" stage of each individual BFE calculation; if a BFE calculation is performed, its corresponding "solvated" stage is simulated for a fixed time of $2 \Lambda$ ns, which incurs a negligible computational cost and is not included in the total allocated computational resource.

\section{Results}
\begin{figure}[ht]
    \centering
    \includegraphics[width=1.0\textwidth]{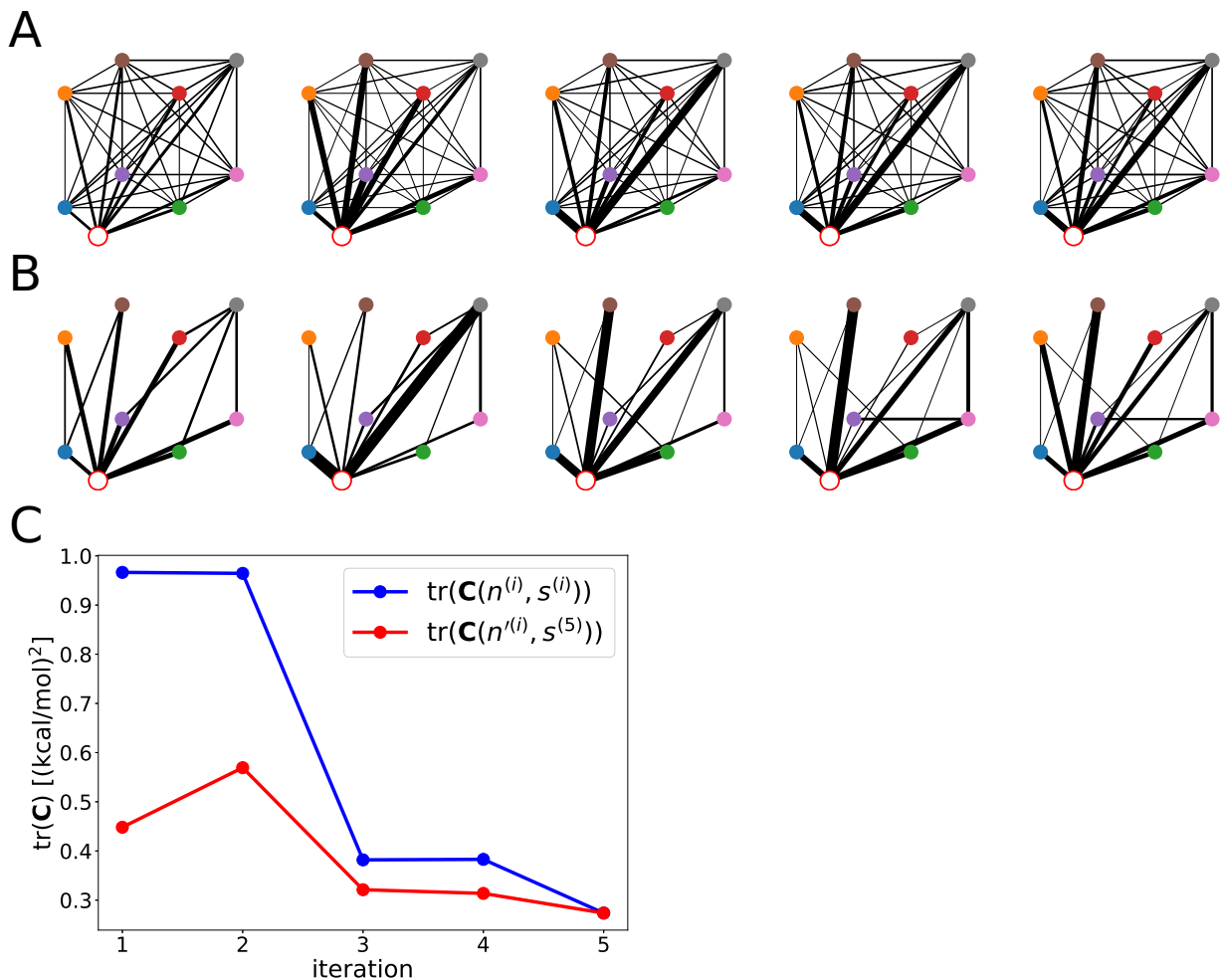}
    \caption{The evolution of a NetBFE calculation of 8 inhibitors against TYK2. \textbf{A}. The estimated statistical fluctuations $s = \{ s_{e} \}$ in each iteration. The width of each edge is proportional to the corresponding $s_{e}$. \textbf{B}. The allocation of $n = \{ n_{e} \}$ in each iteration. The width of each edge is proportional to the corresponding $n_{e}$. In (A) and (B), each filled circle represents an inhibitor, an edge between two filled circles represents a RBFE calculation between the two corresponding inhibitors, and an edge between the large empty circle and a filled circle represents an ABFE calculation for the corresponding inhibitor. \textbf{C}. The total variance $\mathrm{tr}(\mathbf{C}) \equiv$ tr$(\mathbf{C}(n^{(i)}, s^{(i)}))$ according to the estimate of $s^{(i)} = \{ s_e^{(i)} \}$ and the allocation $ n^{(i)} = \{n_e^{(i)} \}$ in each iteration $i$. Also shown is expected total variance tr$(\mathbf{C}(n'^{(i)}, s^{(5)}))$ resulting from the normalized allocation $n'^{(i)} = \sum_e n_e^{(5)} (\sum_e n_e^{(i)})^{-1} n^{(i)}$ in the $i$th iteration if $\{ s_{e}^{(i=5)} \}$ from the last iteration is the true statistical fluctuation; the convergence of $\mathrm{tr}(\mathbf{C}(n'^{(i)}, s^{(5)}))$ to $\mathrm{tr}(\mathbf{C}(n^{(5)}, s^{(5)}))$ demonstrates that NetBFE approaches the A-optimal allocation via the iterative updates.}
    \label{fig:network-iteration}
\end{figure}

We first illustrate the iterative allocation in NetBFE consisting of both ABFE and RBFE calculations, using the small example of eight inhibitors against the protein kinase TYK2. No experimental binding free energies were included in the allocation and analysis, {\it i.e.} $\delta_i^{-1} = 0$ in Eq.~\ref{eq:Fisher}. The estimated fluctuations $\{ s_{e} \}$  and the corresponding A-optimal allocation of $\{ n_{e} \}$ in each iteration are shown in Fig.~\ref{fig:network-iteration}A and Fig.~\ref{fig:network-iteration}B, respectively. The total variance $\mathrm{tr}(\mathbf{C}(n^{(i)}, s^{(i)}))$ decreases with increasing iterations.  

How well does this iterative procedure approach the true A-optimal allocation? If we assume that the estimate of $\{ s_e^{(i=5)} \}$ after the last iteration represents the true statistical fluctuations, $\mathrm{tr}(\mathbf{C}(n^{(5)}, s^{(5)}))$ approximates the variance of the A-optimal allocation.  We measure how far the allocation $n^{(i)}$ in the $i$'th iteration---normalized to $n'^{(i)} \equiv \sum_e n_e^{(5)} (\sum_e n_e^{(i)})^{-1} n^{(i)}$ so that its total number of samples is the same as that in the last iteration---is from the this approximate A-optimal by comparing $\mathrm{tr}(\mathbf{C}(n'^{(i)}, s^{(5)}))$ to $\mathrm{tr}(\mathbf{C}((n^{(5)}, s^{(5)}))$.  The former indeed approaches the latter (Fig.~\ref{fig:network-iteration}C) after only three iterations, suggesting that the iterative allocation is converging quickly to the true A-optimal. 

Because ABFE converges much slower than RBFE (on average ABFE has a much higher statistical fluctuation than RBFE: $\langle \{ s_{i} \} \rangle_{\mathrm{ABFE}}/\langle \{ s_{ij} \} \rangle_{\mathrm{RBFE}}$ $\approx$ 6.2), most of the computations (approximately 81$\%$ of total resources) are allocated to ABFE in order to improve the precision of the absolute values of the computed binding free energies. 
\clearpage

Often it is unnecessary to predict the absolute values of the binding free energies, such as in the case when the values for some similar molecules have been experimentally determined, thus the values for the molecules of interest can be determined by predicting the differences between the molecules, or in the case of ranking the molecules by their binding affinities, such that only their relative differences will suffice.  In such cases ABFE calculations may be omitted to avoid the most expensive part of the computations, as noted above.  In the following, we consider NetBFE with only RBFE calculations.  

\begin{figure}[ht]
    \centering
	\includegraphics[width=1.0\linewidth]{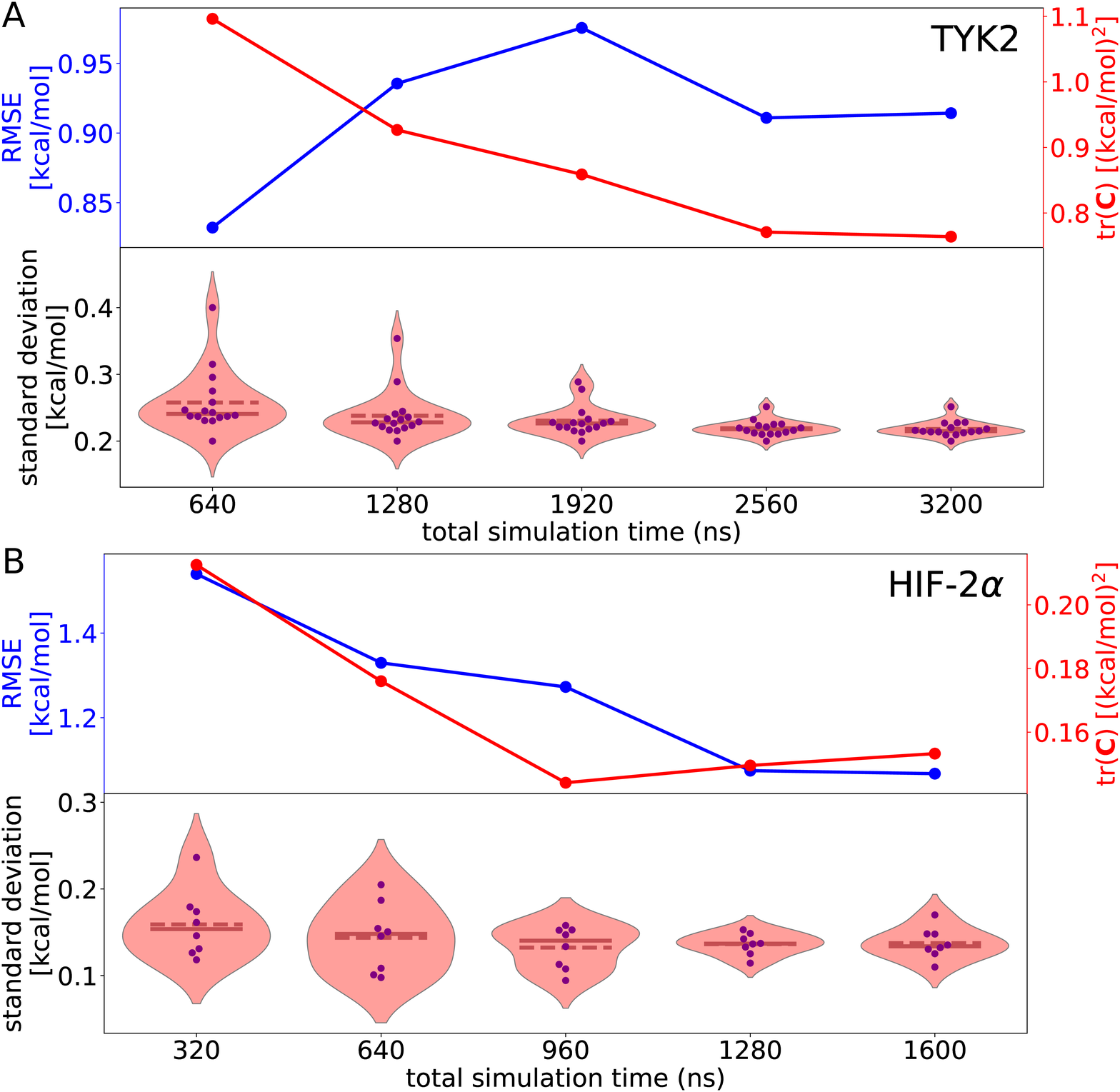}
	\caption{The progression of precision and accuracy in NetBFE iterations for \textbf{A}. 16 inhibitors against TYK2 and \textbf{B}. 8 inhibitors against HIF-2$\alpha$. The upper panels show the total variance tr$(\mathbf{C})$ (red) and the root mean square error (RMSE) between NetBFE predictions and experimental values of the individual binding free energies (blue).  The lower panels show the distributions of the standard deviations in the predicted binding free energies, computed for each inhibitor as the square root of the corresponding diagonal element in $\mathbf{C}$; the dashed and solid lines represent the average and median values, respectively, of the distributions. }
	\label{fig:NetBFE-std-dist-iteration}
\end{figure}

We used NetBFE to compute the binding free energies of 16 inhibitors against TYK2~\cite{Liang2013JMC,Wang2015JACS} and 8 inhibitors against HIF-2$\alpha$~\cite{Scheuermann2015JMC}. These molecules are depicted in Fig.~\ref{fig:mols} and listed in the Supplementary Materials. We first characterized the progression of statistical errors after each iteration of extending the total simulation time (Fig.~\ref{fig:NetBFE-std-dist-iteration}). The standard deviations and the total variance by and large decreases with increasing total simulation time.  Moreover, the distribution of the standard errors becomes narrower, indicating that most of the free energy results have converged to similar precision.
\begin{figure}[ht]	
	\centering
	\includegraphics[width=0.80
	\linewidth]{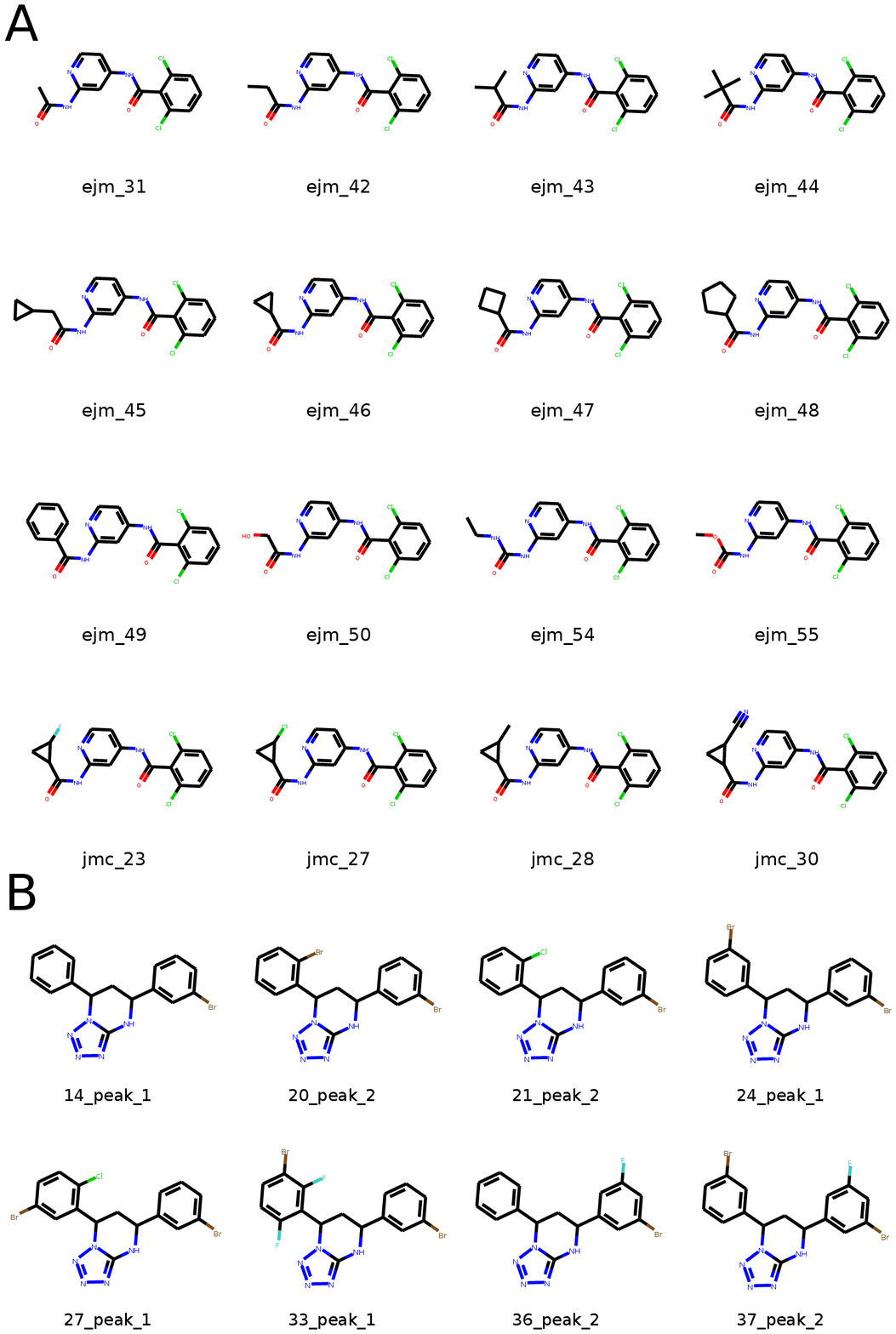}
	\caption{2D structures for the inhibitors studied in this work. \textbf{A}. 16 inhibitors against TYK2; \textbf{B}. 8 inhibitors against HIF-2$\alpha$.}
	\label{fig:mols}
\end{figure}
In the case of HIF-2$\alpha$ inhibitors, the agreement between the experimental binding free energies and their predicted values improves at each iteration (Fig.~\ref{fig:NetBFE-std-dist-iteration}), indicating that the improvement in precision is accompanied by the improvement in accuracy; for TYK2 inhibitors, the accuracy is quite good (RMSE < 1 kcal/mol) in all 5 iterations.  Of course, accuracy in the binding free energy calculations depends on other factors such as the force field parametrization and the consistency between the computational setup and the experimental conditions. It remains unknown whether in other protein-ligand combinations or in other force fields the accuracy of the predicted binding free energies improves with the statistical precision. 
%\clearpage

\begin{figure}[ht]
    \centering
	\includegraphics[width=1.0\linewidth]{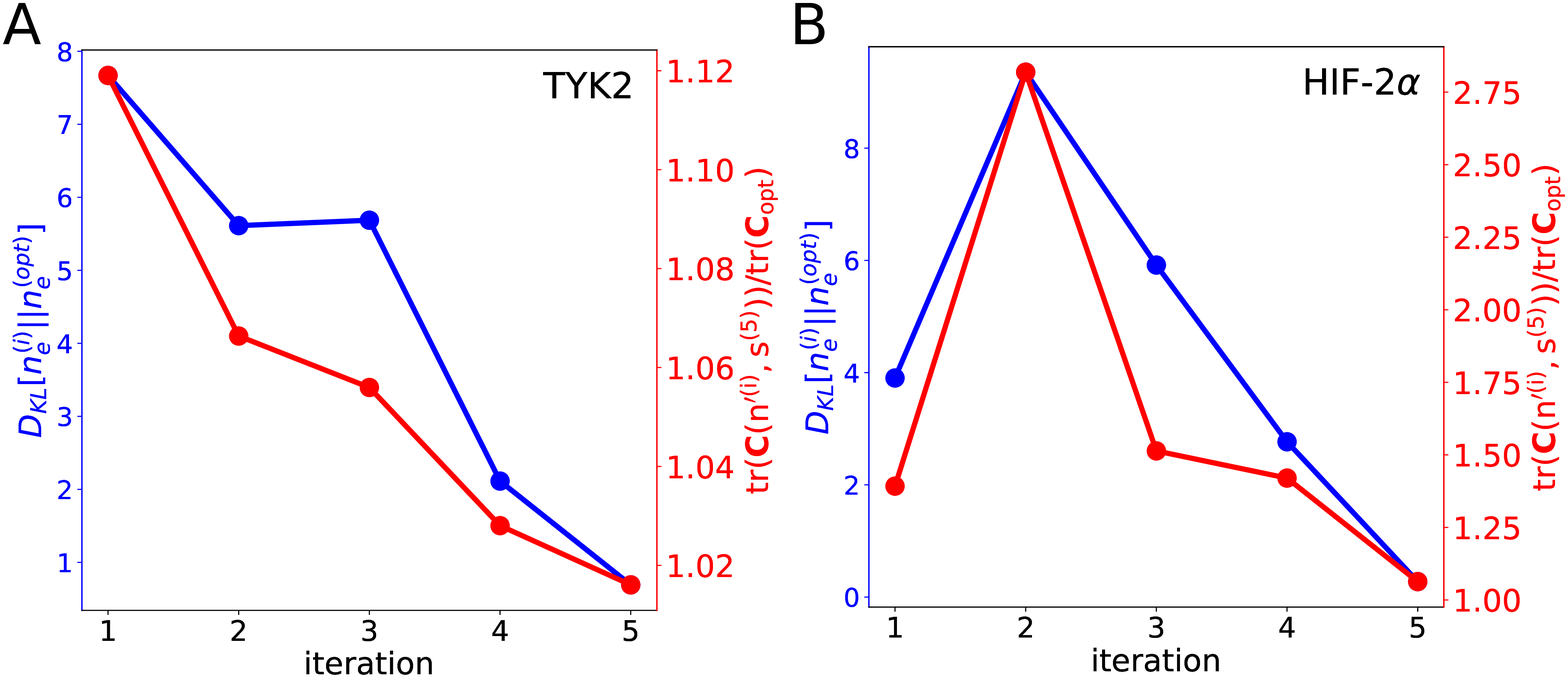}
	\caption{The deviation of allocation in each iteration from the optimal allocation according to the final estimate of $\{ s_e \}$, $D_{\mathrm{KL}}[ n_e^{(i)}||n_e^{(\mathrm{opt})} ]$, and the corresponding ratio tr$(\mathbf{C}(n'^{(i)}, s^{(5)}))$/tr$(\mathbf{C}_{\mathrm{opt}})$.  }
	\label{fig:DKL_RMSD_sij}
\end{figure}

In each iteration of the NetBFE calculation, the allocations $\{ n_{e} \}$ are optimized according to the estimate of statistical fluctuations $\{ s_{e} \}$ based on the results of the previous iteration, thus both $\{ s_{e} \}$ and $\{ n_{e} \}$ change from one iteration to the next.  We measure the difference between two allocations of $\{ n_e \}$ and $\{ n^\ast_e \}$ by the Kullback-Leibler (KL) divergence:
\begin{equation}
    D_{\mathrm{KL}}[n_e||n^\ast_e] = \sum_e f_e \ln \frac{f_e}{f^\ast_e}
\label{eq:KL}
\end{equation}
where $f_e \equiv n_e/\sum_{e'} n_{e'}$ and similarly $f^\ast_e$ are the fractional allocations. In the two cases above, $\{ n_e \}$ approach their limiting values with the progression of iterations (Fig.~\ref{fig:DKL_RMSD_sij}): the KL-divergence between $\{ n_e^{(i)}\}$ and $\{ n_e^{(\mathrm{opt})} \}$ decreases toward 0.69 for TYK2 and 0.28 for HIF-2$\alpha$ with each iteration, where $\{ n_e^{(\mathrm{opt})} \}$ is the the A-optimal allocation based on $s^{(i=5)}$ from the last iteration, assuming, as in the above, that the estimate of $s^{(i=5)}$ from the last iteration is a good approximation to the true statistical fluctuations.  Taking the A-optimal variance to be $\mathbf{C}_{\mathrm{opt}} \approx \mathbf{C}( n^{(\mathrm{opt})}, s^{(5)} )$, $\mathrm{tr}(\mathbf{C}(n'^{(i)}, s^{(5)})/\mathrm{tr}(\mathbf{C}_{\mathrm{opt}})$ rapidly approaches one with each iteration for both TYK2 and HIF-2$\alpha$ examples (as above, $n'^{(i)} = \sum_e n_e^{(5)}(\sum_e n_e^{(i)})^{-1} n^{(i)}$ is the normalized allocation in iteration $i$). Note that $n^{(i)}$ is determined on a 2-connected sub-network (see Section~\ref{sec:subgraph}) while $n^{(\mathrm{opt})}$ is determined on the full network.  These observations suggest that the iterations in NetBFE using a two-connected sub-network result in near-optimal allocations.
\clearpage

\begin{figure}[ht]
    \centering
    \includegraphics[width=1.0\linewidth]{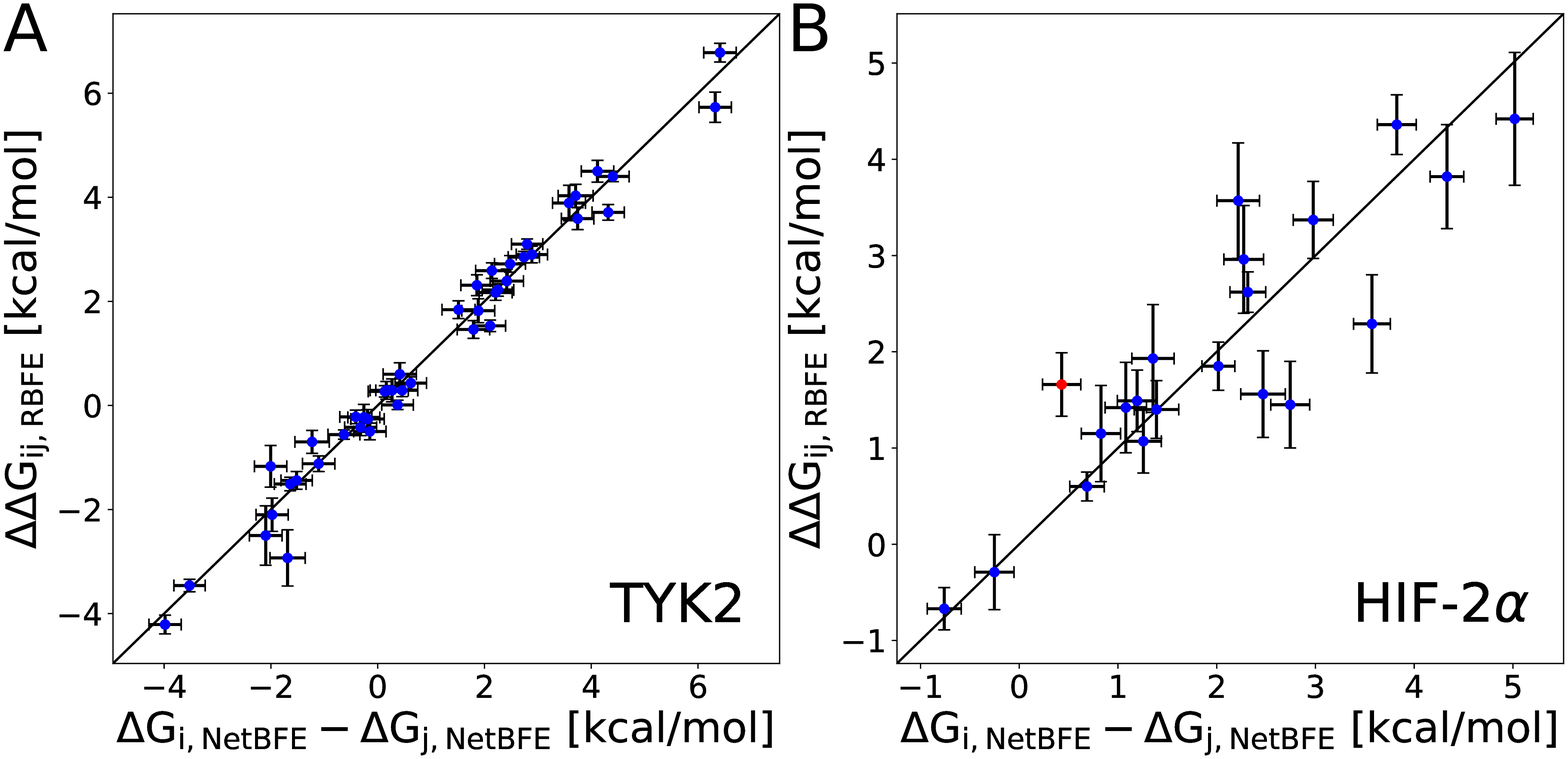}
    \caption{Consistency check between individual RBFE calculations and maximum likelihood estimates of NetBFE results for TYK2 and HIF-2$\alpha$ calculations.  The red point indicates a potentially problematic RBFE calculation inconsistent with NetBFE estimates (see text).}
    \label{fig:RBFE-vs-MLE}
\end{figure}

A simple consistency check can help identify problematic RBFE calculations by comparing the $\Delta\Delta G_{ij, \mathrm{RBFE}}$ in the RBFE calculation against the difference $\Delta G_{i,\mathrm{NetBFE}} - \Delta G_{j,\mathrm{NetBFE}}$, where $\Delta G_{i,\mathrm{NetBFE}}$ and $\Delta G_{j,\mathrm{NetBFE}}$ are derived by the maximum likelihood estimator of Eq.~\ref{eq:MLE}.  Fig.~\ref{fig:RBFE-vs-MLE} shows such comparison for the NetBFE calculations of TYK2 and HIF-2$\alpha$ inhibitors; points significantly deviating from the diagonal (e.g. points for which $|\Delta\Delta G_{ij,\mathrm{RBFE}} - (\Delta G_{i,\mathrm{NetBFE}} - \Delta G_{j,\mathrm{NetBFE}})| \geq 3\sqrt{\sigma_{ij}^2 + \sigma_i^2 + \sigma_j^2} $) indicate potentially problematic RBFE calculations. 
\clearpage

\begin{figure}[ht]
    \centering
    \includegraphics[width=0.9\linewidth]{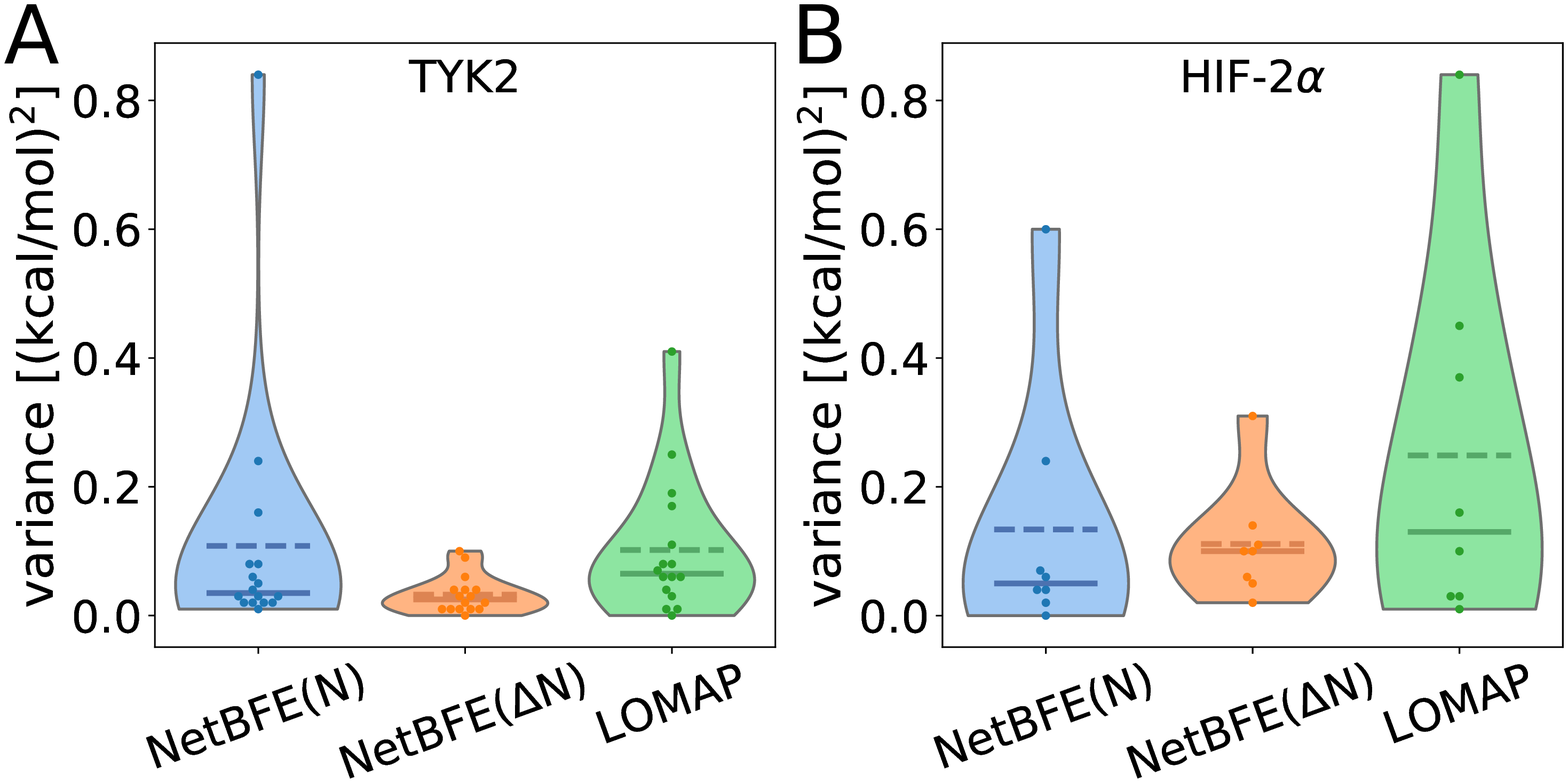}
    \caption{The distributions of variances in the estimated binding free energies (dashed line: mean value of variance distribution; solid line: median value of variance distribution) of NetBFE($N$), NetBFE($\Delta N$) and LOMAP methods for the TYK2 and HIF-2$\alpha$ examples. The variances are computed from 3 replicates of binding free energy calculations.}
    \label{fig:NetBFE-var}
\end{figure}

Having demonstrated above that the iterative procedure in NetBFE allocates the computational resources approximately according to the A-optimal, we next investigate whether such allocations yield improved precision in the computed binding free energies.  We performed $g=3$ independent replicates of NetBFE calculations of the TYK2 and HIF-2$\alpha$ examples, initializing each replicate with the same binding poses but different random number seeds in equilibration and alchemical simulations, and computed the statistical variances in the predicted binding free energies across the replicates. In comparison, we computed the same sets of binding free energies---and their statistical variances from $g=3$ independent replicates---by allocating the same total computational cost evenly to pairs of RBFE calculations selected by a previously published method, LOMAP~\cite{Liu2013JCAMD}. Fig.~\ref{fig:NetBFE-var} shows the distributions of variances for the two NetBFE allocation methods (NetBFE($N$) and NetBFE($\Delta N$), explained in Section~\ref{sec:subgraph}) and the LOMAP allocation method. The mean and median values of the variance distributions are listed in Table~\ref{tab:NetBFE-LOMAP-var}. The NetBFE allocations indeed significantly outperform the LOMAP allocations in terms of the statistical precision in the predicted binding free energies. The reduction in the statistical variance by approximately a factor of two is consistent with the previous results from simulated data~\cite{Xu2019JCIM}. 

\begin{table*}
	\setlength\tabcolsep{3 pt}
	\newcommand{\rb}[1]{\raisebox{1.5ex}[0pt]{#1}}
	\caption{The mean and median values of the variances distributions in the estimated binding free energies of NetBFE($N$), NetBFE($\Delta N$) and LOMAP schemes for the TYK2 and HIF-2$\alpha$ examples, corresponding to the dashed and solid lines in Fig.~\ref{fig:NetBFE-var}. The error bars for these values are computed from bootstrapping. The variances are in the units of (kcal/mol)$^2$.}
	\label{tab:NetBFE-LOMAP-var}
	\begin{tabular}{lcccccccccc}
		\hline
		&\multicolumn{2}{c}{TYK2}&&\multicolumn{2}{c}{HIF-2$\alpha$}&\\
		\cline{2-3}\cline{5-6}
		\rb{methods}&mean&median&&mean&median&\\
		\hline
		NetBFE($N$)&0.11 $\pm$ 0.05&0.04 $\pm$ 0.02&&0.13 $\pm$ 0.07& 0.05 $\pm$ 0.06 \\
		NetBFE($\Delta N$)&0.03 $\pm$ 0.01&0.03 $\pm$ 0.01&&0.11 $\pm$ 0.03& 0.10 $\pm$ 0.02 \\
		LOMAP&0.10 $\pm$ 0.03&0.07 $\pm$ 0.02&&0.25 $\pm$ 0.10& 0.13 $\pm$ 0.13 \\
		\hline
	\end{tabular}
\end{table*}
%\clearpage

Reference binding free energy values from experimental measurements for a subset of the molecules, when available, may be incorporated in the NetBFE calculations by supplying $\tilde{x}_i$ and $\delta_i^{-1} > 0$ in Eq.~\ref{eq:Fisher},~\ref{eq:z}.  Here we tested the allocation and the corresponding precision and accuracy in NetBFE calculations that include different numbers of reference values.  We performed four independent NetBFE calculations for the 16 inhibitors against TYK2, supplying reference experimental values for $k=0, 1, 2, 4$ molecules, respectively.  

\begin{figure}[ht]
    \centering
	\includegraphics[width=1.0\linewidth]{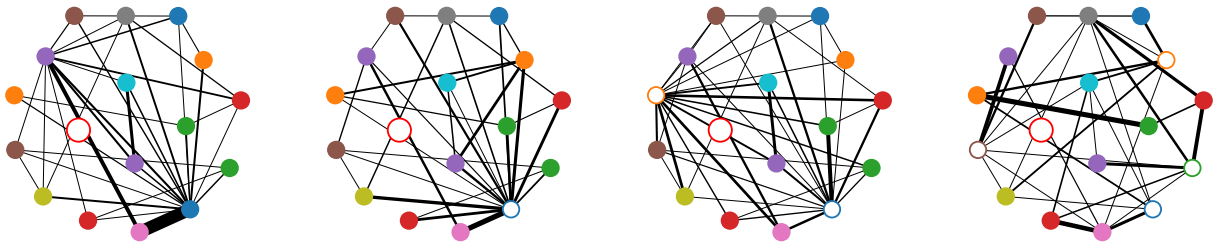}
	\caption{The network of optimal allocations when including $k=0, 1, 2, 4$ reference experimental binding free energies.  The small empty circles in the graph represent the ligands whose experimental binding free energies are used as reference values. The disconnected apo node (the large empty circle) indicates that there are no ABFE calculations.}
	\label{fig:NetBFE-with-exp:allocations} 
\end{figure}

\begin{figure}[ht]
    \centering
	\includegraphics[width=1.0\linewidth]{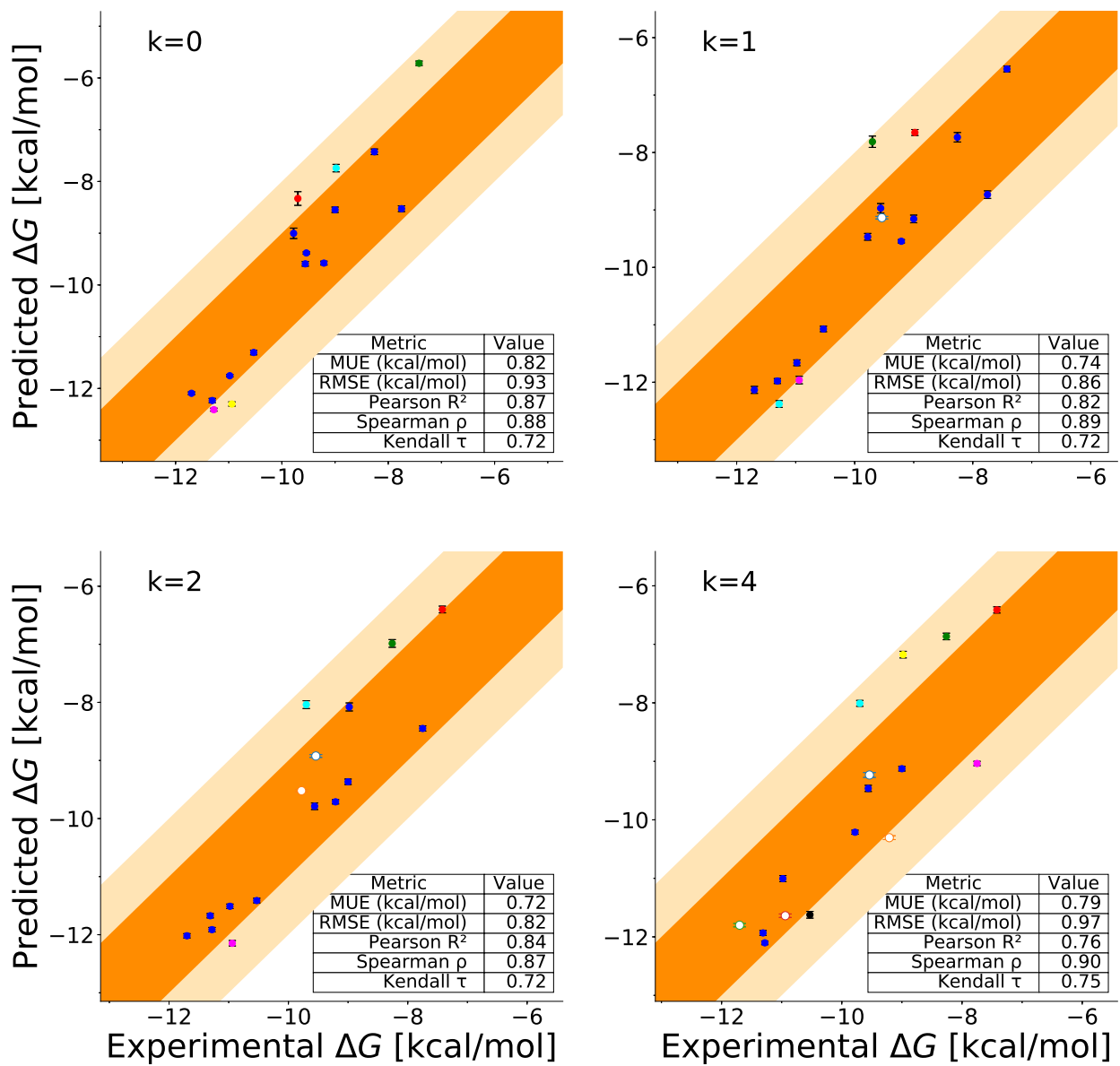}
	\caption{\label{fig:NetBFE-with-exp:metric} The predicted binding free energies by NetBFE versus experimental values when including $k=0, 1, 2, 4$ reference experimental binding free energies. Here, the predicted binding free energies by NetBFE are shifted to minimize the difference between the means of the predicted and the experimental values. The error bars in the predicted values are the square roots of the diagonal elements of the covariance matrix in the NetBFE calculations. The molecules with reference values are represented by unfilled circles; their predicted binding free energies are included in the calculation of MUE, RMSE, and the correlation coefficients.}
\end{figure}

The optimal allocations at the end of five iterations are shown in Fig.~\ref{fig:NetBFE-with-exp:allocations}.  When there are a small number of reference molecules ($k=1, 2$), they serve as network hubs, in that there is a RBFE edge connecting them to every other molecule.  As we increase the number of reference molecules ($k=4$), however, only a partial subset of the molecules are connected to each reference molecule for which RBFE calculations are quick to converge. The predicted binding free energies are compared to the experimental values in Fig.~\ref{fig:NetBFE-with-exp:metric}. The precision and accuracy of the NetBFE calculations are summarized in Table~\ref{tab:NetBFE-with-k-references}.  In this case, including more reference values does not significantly improve either the precision or the accuracy.

\begin{table}[t]
    \centering
    \begin{tabular}{rcccc}
         k & 0 & 1 & 2 & 4 \\
         \hline
         $m^{-1}\mathrm{tr}(\mathbf{C})$  & 0.0030 $\pm$ 0.0011 & 0.0040 $\pm$ 0.0006 & 0.0020 $\pm$ 0.0003 & 0.0020 $\pm$ 0.0002 \\
         MUE & 0.82 $\pm$ 0.12 & 0.74 $\pm$ 0.11 & 0.72 $\pm$ 0.10 & 0.79 $\pm$ 0.14 \\
         RMSE & 0.93 $\pm$ 0.11 & 0.86 $\pm$ 0.12 & 0.82 $\pm$ 0.11 & 0.97 $\pm$ 0.13 \\
         Pearson $R^2$ & 0.87 $\pm$ 0.07 & 0.82 $\pm$ 0.09 & 0.84 $\pm$ 0.07 & 0.76 $\pm$ 0.11 \\
         Spearman $\rho$ & 0.88 $\pm$ 0.09 & 0.89 $\pm$ 0.10 & 0.87 $\pm$ 0.10 & 0.90 $\pm$ 0.08 \\
         Kendall $\tau$ & 0.72 $\pm$ 0.11 & 0.72 $\pm$ 0.12 & 0.72 $\pm$ 0.12 & 0.75 $\pm$ 0.10
    \end{tabular}
    \caption{The precision and accuracy of NetBFE calculations for $m=16$ TYK2 inhibitors when reference experimental values are included for $k=0,1,2,4$ molecules.  The precision is measured by the average statistical variance in the free energy estimates (in (kcal/mol)$^2$).  The accuracy is measured by the mean unsigned error (MUE, in kcal/mol) and root mean square error (RMSE, in kcal/mol) from the experimental values and by the correlations (Pearson $R^2$, Spearman $\rho$, and Kendall $\tau$) with the experimental values.  The error bars in the metrics are computed from bootstrapping.}
    \label{tab:NetBFE-with-k-references}
\end{table}

\clearpage

\section{Conclusions}
In this work, we have developed a new method to improve the overall statistical precision of binding free energy calculations for multiple molecules against a common biomolecular target.  NetBFE (Network Binding Free Energy) implements and extends a previous theoretical work from our group on A-optimal experimental design for measuring pairwise differences, making it practical for realistic binding free energy calculations.  We have demonstrated that, at a fixed total computational cost, NetBFE can reduce the statistical errors in BFE calculations by approximately two-fold.  The improved precision will increase the confidence in the computational ranking of candidate drug molecules, which is instrumental to prioritizing molecules for synthesis in the design-make-test cycle.  In addition, as binding free energy calculations are increasingly used for validating and benchmarking force field models and free energy methods\cite{Schindler2020JCIM, Wang2015JACS, Mey2020LJCMS, David2021arXiv}, the reduced statistical errors afforded by NetBFE will make the benchmark assessments more reliable and thus accelerate the improvements in {\it in silico} binding assays.
\clearpage

\bibliography{bibliography}

\end{document}